# A Downstream Crosstalk Channel Estimation Method for Mix of Legacy and Vectoring-Enabled VDSL

Mehdi Mohseni, Wonjong Rhee, Georgios Ginis
{mmohseni@assia-inc.com, wrhee@snu.ac.kr, gginis@gmail.com}

*Abstract*—With the latest technology of vectoring, DSL data rates in the order of 100Mbps have become a reality that is under field deployment. The key is to cancel crosstalk from other lines, which is also known as multiuser MIMO cancellation for wireless communications. During the DSL system upgrade phase of field deployment, mix of legacy and vectoring-enabled VDSL lines is inevitable and a channel estimation solution for the entire mix is needed before vectoring can be enforced. This paper describes a practical method for crosstalk channel estimation for downstream vectoring, assuming that a vectoring-enabled DSLAM forces DMT symbol-level timing to be aligned for all of the lines, but also assuming that the location of synch symbols are aligned only among vectoring-enabled lines. Each vectoring-enabled receiver is capable of reporting error samples to vectoring-DSLAM. The estimation method is not only practical, but also matches the performance of Maximum-Likelihood estimator for the selected training sequences.

## I. INTRODUCTION

A major obstacle for delivering very high-speed communications services over copper has been crosstalk interference among neighboring pairs. With the latest DSL technology advances, however, data rates in the order of 100 Mbps are now a reality. Network infrastructure and consumer hardware can be designed to perform crosstalk cancellation [1][2] of VDSL2 [3] signals over twisted pairs, and thereby enable dedicated speeds that were previously considered possible only with unshared fiber. This crosstalk-cancelling technology is called "vectoring," and is defined in the ITU-T recommendation G.993.5 (also known as G.vector) [4].

Vectoring performs multiuser signal coordination at the DSLAM (DSL Access Multiplexer; network operator side equipment) for both upstream and downstream. Crosstalk cancellation is applied to upstream, and crosstalk pre-coder is applied to downstream [1][2][4]. The concept and theory have been known, and vectoring has been successfully implemented for demo and early commercial deployments. The vectoring technology is a major milestone for broadband access and opens many opportunities for new services. Vectored DSL allows super-fast connections comparable to those of fiber-to-the-home installations. However, vectored DSL often has the advantage of being dramatically more economical than any fiber-to-the-home investment.

A network upgrade to vectored DSL cannot take place overnight. There are several practical reasons why vectored DSLs are expected to share binders with non-vectored DSLs for a long time. Legacy DSL equipment at the customer site, Customer Premise Equipment (CPE), cannot be instantly replaced with vectored DSL equipment. A remote firmware upgrade is a preferred option, but may still fail for some of the lines. In this work, a single vectored DSLAM connected with a mix of legacy and vectoring-enabled CPEs is considered (see Fig. 1.). Additional to CPEs, legacy and vectored DSLAM hardware may share a cabinet and lead to cases of "coexistence" of vectored and non-vectored DSLs in the same binder. This deployment scenario is not in the scope of this study[1].

To perform vectoring, accurate channel estimation is needed. It is less difficult for upstream, because all synch symbols of legacy and vectoring-enabled lines are known to the DSLAM. The knowledge can be used for crosstalk channel estimation. For downstream, however, feedback from CPEs is required for performing crosstalk channel estimation. Legacy VDSL CPEs cannot provide such information, but vectoring-enabled CPEs have been designed to supply the information. Then, the goal is to estimate crosstalk channels from *all* DSL lines to the vectoring-enabled lines, because only vectoring-enabled lines can utilize pre-coding that is analogous to crosstalk cancelation of upstream.

To maintain orthogonality of DMT symbols, timing of legacy and vectoring-enabled lines need to meet the requirements explained in Fig. 2. DMT symbol timing of all lines, including vectored and legacy lines, must be aligned and this can be enforced by DSLAM during handshaking. The location of synch symbols is required to be aligned among the vectoring-enabled lines, but not aligned for the legacy lines. In this way, a full design flexibility over synch symbols of vectoring-enabled lines can be obtained and a full randomness can be maintained over legacy lines whose synch symbols follow legacy system's design. The assumptions listed so far for downstream crosstalk channel estimation are the working assumptions for recent implementations of vectored DSL [6].

Before moving to the channel estimation section, it is worth mentioning one important issue for realizing vectoring. In a real deployment, often CPEs are turned off and on by the user. Furthermore, sometimes a CPE resets itself to recover from an uncorrectable error. Typically, off-and-on was not a serious issue for non-vectored DSL, but it can cause a serious problem for vectored DSL. An unexpected turning on event of a DSL can cause an abrupt change in crosstalk noise for other DSL(s) that have been running, and the change can be large enough to cause the affected DSL(s) to experience a very large error rate. Consequently, the error event can trigger the affected DSLs to reset and make the entire DSL group unstable. To prevent such an instability, a good solution is to force any turning-on DSL to transmit synch symbols only until channel estimation is completed and pre-coding and cancelation are ready to run with the new line included. This idea, however, is applicable only to vectoring-enabled lines and legacy lines need to be tuned to transmit at low power level to protect other DSL lines. The low power requirement for legacy DSL means legacy lines will not be allowed to achieve a very high data rate, but rather assigned of legacy low-rate service. The solution is not perfect, but greatly useful when operators are slowly migrating millions of legacy lines to vectored VDSL with high rate services.

---

[1] Technicians may attempt to use two binders with minimum crosstalk (due to physical separation) – one for vectored DSLAM and the other for legacy DSLAM. Typically, however, such a re-assignment of copper pairs is not an easy task to complete.

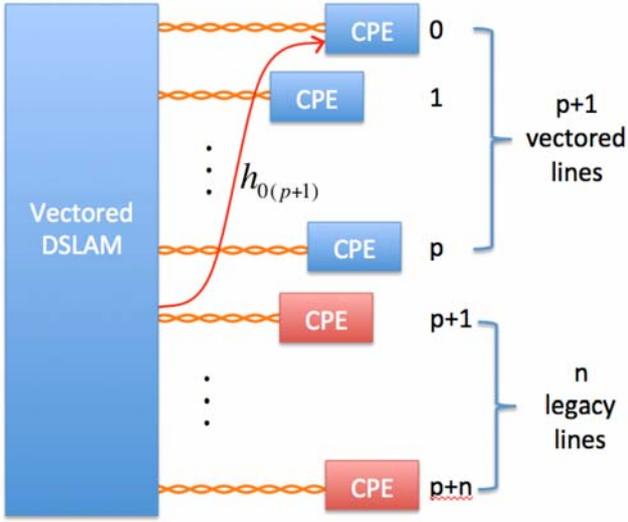

Fig. 1. Co-existence of vectoring-enabled VDSL2 CPE's and legacy VDSL2 CPE's. The copper lines are densely packed in a trunk and the crosstalk channels, such as $h_{0(p+1)}$ that represents the channel between $(p+1)$'th transmitter and 0'th receiver, can be either strong or weak depending on the electro-magnetic coupling between the two copper pairs.

## II. SYSTEM MODEL AND CHANNEL ESTIMATION PROBLEM

Because of the time alignment of DMT-symbols, orthogonality is preserved not only for the intended signal but also for crosstalk signals. Therefore, each subchannel can be analyzed independently, and the subchannel index is omitted throughout this paper. Then, the system in Fig. 1. can be modeled as following for a single subchannel.

$$y_k(t) = h_{kk}\sqrt{s_k}\, x_k(t) + \sum_{j \neq k} h_{kj}\sqrt{s_j}\, x_j(t) + z_k(t)$$

$k$ is line index, $t$ is DMT symbol index, and $y_k(t)$ and $x_k(t)$ are signals received at and transmitted for CPE $k$ at DMT symbol $t$, respectively. $z_k(t)$ is i.i.d. Gaussian noise with power $\sigma_k^2$. $h_{kk}$ is line $k$'s own channel and $h_{kj}$ is crosstalk channel between transmitter $j$ and receiver $k$. The channel variations are assumed to be negligible, and hence $h_{kk}$ and $h_{kj}$ are independent of $t$. The power adjustment factor $s_k$ is assumed to depend on line index $k$ only, and it can be absorbed into channel for the mathematical representation of the problem.

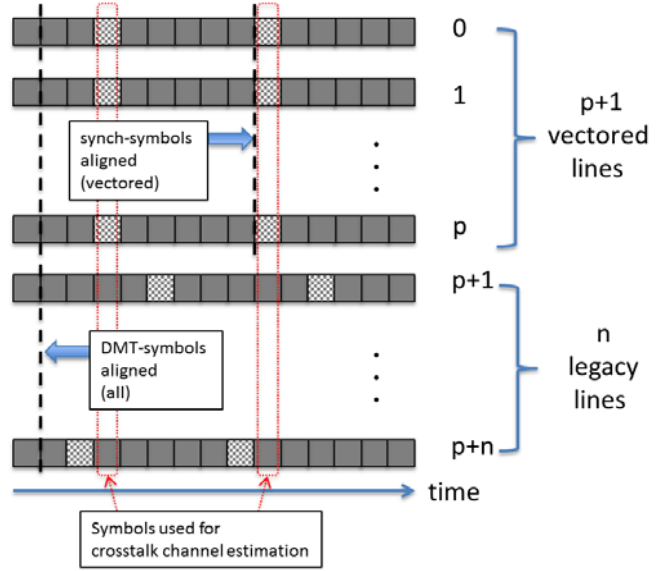

Fig. 2. Time diagram of DMT symbols. Checker-shaped ones are synch symbols and solid ones are data symbols. Vectored-DSLAM forces DMT symbol-level timing to be aligned for all lines, but the location of synch symbols are aligned only among vectored lines.

$$y_k(t) = h_{kk}\, x_k(t) + \sum_{j \neq k} h_{kj}\, x_j(t) + z_k(t) \quad (1)$$

The direct channel $h_{kk}$ is estimated in a conventional way for both legacy and vectored lines, and assumed to be known. The problem of crosstalk channel estimation is to estimate $h_{kj}$ ($k = 0, 1, \cdots, p$ and $j \neq k$,), such that the crosstalk channel information can be used for Vectored-DSLAM's pre-coding. For the estimation, vectoring-enabled CPE's support feedback of the below error value for the synch-symbols with known transmitted training data.

$$\begin{aligned} e_k(t) &= \frac{y_k(t)}{h_{kk}} - x_k(t) \\ &= \sum_{j \neq k} \frac{h_{kj}}{h_{kk}} x_j(t) + \frac{z_k(t)}{h_{kk}}, \quad k = 0, 1, \cdots p \end{aligned} \quad (2)$$

Note that $e_k(t)$ is available only for vectoring-enabled lines. Because $h_{kk}$ is known, the estimation problem is equivalent to estimating the following.

$$\tilde{h}_{kj} = \frac{h_{kj}}{h_{kk}} \quad (3)$$

Without loss of generality, only $k = 0$ is considered in this paper and the crosstalk channels of the other vectored lines can be estimated in the same way. Assuming that $m$ of DMT synch-symbols are used for the crosstalk channel estimation, the following error vector can be constructed for the chosen training sequence $V$ for vectoring-enabled lines and for the user data sequence $L$ for legacy lines.

$$\underline{e}_0 = [e_0(1)\ e_0(2)\ \cdots\ e_0(m)]^T$$
$$= V\underline{h}_{0v} + L\underline{h}_{0l} + \underline{z}_0, \quad (4)$$

where

$$\underline{h}_{0v} = \begin{bmatrix} \tilde{h}_{01} \\ \tilde{h}_{02} \\ \vdots \\ \tilde{h}_{0p} \end{bmatrix}, \underline{h}_{0l} = \begin{bmatrix} \tilde{h}_{0(p+1)} \\ \tilde{h}_{0(p+2)} \\ \vdots \\ \tilde{h}_{0(p+n)} \end{bmatrix}, and\ \underline{z}_0 = \frac{1}{h_{00}} \times \begin{bmatrix} z_0(1) \\ z_0(2) \\ \vdots \\ z_0(m) \end{bmatrix}$$

are $p \times 1$, $n \times 1$ and $m \times 1$ vectors, respectively. $\underline{h}_{0v}$ represents the crosstalk channels between transmitters of vectoring-enabled lines and receiver 0. $\underline{h}_{0l}$ represents the crosstalk channels between transmitters of legacy lines and receiver 0. $\underline{z}_0$ is the vector of i.i.d. Gaussian noise samples for the $m$ DMT synch-symbols. And,

$$V = \begin{bmatrix} \underline{x}_1 & \underline{x}_2 & \cdots & \underline{x}_p \end{bmatrix} = \begin{bmatrix} x_1(1) & x_2(1) & \cdots & x_p(1) \\ x_1(2) & x_2(2) & \cdots & x_p(2) \\ \vdots & \vdots & \ddots & \vdots \\ x_1(m) & x_2(m) & \cdots & x_p(m) \end{bmatrix}$$

is $m \times p$ matrix with each column consisting of $m$ transmitted training symbols to each of the $p$ vectoring-enabled CPE's, and

$$L = \begin{bmatrix} \underline{x}_{p+1} & \underline{x}_{p+2} & \cdots & \underline{x}_{p+n} \end{bmatrix} = \begin{bmatrix} x_{p+1}(1) & x_{p+2}(1) & \cdots & x_{p+n}(1) \\ x_{p+1}(2) & x_{p+2}(2) & \cdots & x_{p+n}(2) \\ \vdots & \vdots & \ddots & \vdots \\ x_{p+1}(m) & x_{p+2}(m) & \cdots & x_{p+n}(m) \end{bmatrix}.$$

is $m \times n$ matrix with each column consisting of $m$ transmitted user data symbols to each of the $n$ legacy CPE's. $L$ is known to DSLAM and can be used for the estimation because the user data is transmitted by DSLAM.

### III. PROPOSED ESTIMATION METHOD

The problem can be summarized as estimating $\underline{h}_{0v}$ and $\underline{h}_{0l}$ for the given $\underline{e}_0$, $V$, and $L$ at DSLAM. While $L$ is user data and cannot be controlled (but known), $V$ is training data that can be designed and chosen. To simplify the estimation process, $V$ is selected as a multiplication of Walsh-Hadamard orthogonal binary sequences (sequences of $\pm 1$) and randomly chosen 4-QAM diagonal matrix as below.

$$V = WA, \quad A = \begin{bmatrix} \alpha_1 & 0 & \cdots & 0 \\ 0 & \alpha_2 & \cdots & 0 \\ \vdots & \vdots & \ddots & \vdots \\ 0 & 0 & \cdots & \alpha_p \end{bmatrix} \quad (5)$$

$W$ is $m \times p$ matrix constructed by choosing the first $p$ columns of an $m \times m$ Hadamard matrix [2] and $\alpha_1, \alpha_2, \cdots, \alpha_p$ are randomly chosen 4-QAM symbols. With this choice of $V$, the columns of $V$ are orthogonal to each other and the following holds.

$$V^H V = 2m I_p \quad (6)$$

$V^H$ is conjugate transpose of $V$ and $I_p$ is an identity matrix of size $p$. The factor 2 above is due to the energy of each 4-QAM symbol.

In the method described herein, the estimation of the normalized crosstalk coupling coefficients from the legacy VDSL2 loops is considered first. Let $U$ be an $m \times (m-p)$ matrix such that $[W\ U]$ forms the original $m \times m$ Hadamard matrix used for constructing $W$. In other words, $U$ consists of the remaining columns of the Hadamard matrix. Since Hadamard matrix is an orthogonal matrix, $U^H W = 0$ and therefore $U^H V = U^H W A = 0$.

Let $Q$ be a projection matrix into the subspace perpendicular to the range of matrix $V$. One choice of $Q$ can be as below:

$$Q = I_m - \frac{1}{2m} V V^H$$

Clearly, $QV = V - \frac{1}{2m} V V^H V = 0$ because of (6), and any vector in the range of $V$ is projected to zero. Let $U$ be an $m \times (m-p)$ matrix such that $[W\ U]$ forms the original $m \times m$ Hadamard matrix used for constructing $W$. In other words, $U$ consists of the remaining columns of the Hadamard matrix. Since Hadamard matrix is an orthogonal matrix, $UU^H + WW^H = mI_m$. Because $VV^H = 2WW^H$ from direct evaluation of (5), the following can be derived.

$$Q = I_m - \frac{1}{2m} VV^H = I_m - \frac{1}{m} WW^H = \frac{1}{m} UU^H$$

Therefore, it can be seen that $U^H$ also projects any vector in the range of $V$ to zero. Then, $U^H$ can be applied to $\underline{e}_0$ to filter out signals related to $\underline{h}_{0v}$ and the below follows.

$$U^H \underline{e}_0 = U^H L \underline{h}_{0l} + U^H \underline{z}_0 \quad (7)$$

Now pseudo-inverse can be applied to (7) to find the maximum-likelihood estimation of $\underline{h}_{0l}$.

$$\widehat{\underline{h}}_{0l} = (L^H UU^H L)^{-1} L^H UU^H \underline{e}_0 \quad (8)$$

By subtracting (8) from (4) with a proper scaling factor, $\underline{h}_{0v}$ can be estimated.

$$\widehat{\underline{h}}_{0v} = \frac{1}{2m} V^H (\underline{e}_0 - L\widehat{\underline{h}}_{0l})$$
$$= \frac{1}{2m} V^H (I_m - L(L^H UU^H L)^{-1} L^H UU^H) \underline{e}_0 \quad (9)$$

(8) and (9) together forms the crosstalk channel estimations of the proposed method. The estimates depends on $\underline{e}_0$, $V$, $U$ and $L$ only and they are all available at DSLAM.

---

[2] Hadamard matrix does not exist for all $m$. It is assumed that $m$ is sufficiently larger than $p$ and that $m$ is chosen as a number that is power of 2 value without affecting estimating performance.

## IV. OPTIMALITY OF THE PROPOSED ESTIMATION METHOD

The proposed method derives $\underline{h}_{0l}$ first, and the estimate is used to estimate $\underline{h}_{0v}$. The sequential derivation of the two vectors significantly reduces the computational complexity, because the computations are applied to smaller size matrices compared to joint maximum-likelihood estimation of the both. In this section, (8) and (9) are shown to be not only computationally efficient, but also equivalent to the joint maximum-likelihood estimation of $\underline{h}_{0v}$ and $\underline{h}_{0l}$ because of the particular selection of training sequence $V$.

Equation (4) can be re-written as below.

$$\underline{e}_0 = [V\ L]\begin{bmatrix}\underline{h}_{0v}\\\underline{h}_{0l}\end{bmatrix} + \underline{z}_0$$

By applying pseudo-inverse to $\begin{bmatrix}\underline{h}_{0v}\\\underline{h}_{0l}\end{bmatrix}$ together, the joint maximum-likelihood estimates can be expressed as follow.

$$\begin{bmatrix}\hat{\underline{h}}_{0v}^{ML}\\\hat{\underline{h}}_{0l}^{ML}\end{bmatrix} = \left(\begin{bmatrix}V^H\\L^H\end{bmatrix}[V\ L]\right)^{-1}\begin{bmatrix}V^H\\U^H\end{bmatrix}\underline{e}_0$$
$$= \begin{bmatrix}2mI_p & V^HL\\L^HV & L^HL\end{bmatrix}^{-1}\begin{bmatrix}V^H\\U^H\end{bmatrix}\underline{e}_0 \quad (10)$$

For the inverse of 2 by 2 block matrix, the blockwise inversion theory [5] can be applied to derive the following.

$$\begin{bmatrix}\hat{\underline{h}}_{0v}^{ML}\\\hat{\underline{h}}_{0l}^{ML}\end{bmatrix} = \begin{bmatrix}\frac{1}{2m}I_p + \frac{1}{4m^2}V^HLSL^HV & -\frac{1}{2m}V^HLS\\-\frac{1}{2m}SL^HV & S\end{bmatrix}\begin{bmatrix}V^H\\U^H\end{bmatrix}\underline{e}_0 \quad (11)$$

where,

$$S = \left(L^H\left(I_m - \frac{1}{2m}VV^H\right)L\right)^{-1}$$
$$= m(L^HUU^HL)^{-1}$$

because of the way $V$ is chosen. After replacing $S$ with $m(L^HUU^HL)^{-1}$ into (11), the joint maximum likelihood estimates simplify to the following.

$$\begin{bmatrix}\hat{\underline{h}}_{0v}^{ML}\\\hat{\underline{h}}_{0l}^{ML}\end{bmatrix} = \begin{bmatrix}\frac{1}{2m}V^H(I_m - L(L^HUU^HL)^{-1}L^HUU^H)\\(L^HUU^HL)^{-1}L^HUU^H\end{bmatrix}\underline{e}_0$$

These estimates are the same as in (8) and (9), and the result proves that the proposed method can perform as well as the joint maximum-likelihood estimation of $\underline{h}_{0v}$ and $\underline{h}_{0l}$.

Note that (8) and (9) require matrix inversion of an $n \times n$ matrix for each, but (10) requires matrix inversion of a $(p+n) \times (p+n)$ matrix. (10) simplifies to (8) and (9) only because Walsh-Hadamard orthogonal binary sequences were used as expressed in (5).

## V. SIMULATION RESULTS

Simulations were run to validate the analysis. In Fig. 3, receiver SINR of the CPE of interest ($k = 0$) is shown for Walsh-Hadamard code length between 4 and 64. The number of interfering vectored lines is fixed at 4, and the number of interfering legacy (non-vectored) lines is varied to 1, 4 and 10. The crosstalk-channel is assumed to be 6dB weaker from the other vectored lines, and 10dB or 15dB weaker from the legacy lines. Noise power is chosen such that receiver SNR is 30dB without any crosstalk.

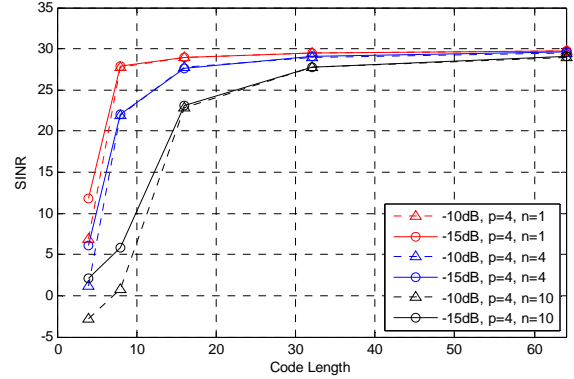

Fig. 3. SINR (Signal to Interference and Noise Ratio) of CPE of interest ($k = 0$) after running channel estimation and using the estimated channel for decoding. Crosstalk channel from legacy lines is either 10dB or 15dB weaker compared to the signal's path.

As can be seen in the figure, SINR is affected by the number of crosstalk lines and the strength of crosstalk channel, but it becomes very close to 30dB when code length is 32 or larger.

## VI. CONCLUSION

In this paper, a practical crosstalk channel estimation method is considered for mix of legacy and vectoring-enabled VDSL that are connected to a single DSLAM. Use of Walsh-Hadamard as the training sequence of vectoring-enabled VDSL is shown to be effective for reducing estimation complexity. With the proposed estimation method, vectoring is shown to achieve almost complete crosstalk cancellation from both vectoring-enabled and legacy groups. The proposed method achieves the performance of maximum-likelihood estimator for the chosen training sequence, and a proof is provided.